\documentclass[aps,preprint]{revtex4}

\begin{document}

\bibliographystyle{prsty}

\draft


\title{Two years later--lessons from vortex dynamics in super media}

\author{ P. Ao }
\address{ Departments of Mechanical Engineering and Physics,
          University of Washington, Seattle, WA 98195, USA }
\date{ October 24, 2006 }


\begin{abstract}
 Vortex dynamics in super media has been one of most intriguing
fields in physics. It is therefore not surprising that it had
troubled great physicists such as L. Landau. 50 years after its
conception, in 1990's the consistent fundamental vortex dynamics
equation emerged after long and serious effort by a group of
researchers associated with Seattle. Nevertheless, this
fundamental equation has been met with resistance. Two years ago
the present author analyzed this situation (cond-mat/0407007).
Five "mistakes" were identified to explain this resistance. Given
the current tremendous interest in vortex dynamics, it would be
desirable to provide a progress report: A survey of literature
reveals that 3 out 5 "mistakes" has in fact been confirmed by
other researchers, not a bad appreciation.
\end{abstract}



\maketitle

\section{Introduction}

To understand the resistance in the acceptance of the fundamental
equation of vortex dynamics in super media, five "mistakes" were
identified two years ago. They are: \\
1) Misuse of the relaxation time approximation; \\
2) Mixing with another effect; \\
3) Double counting the same topological effect; \\
4) No extra Berry phase contribution at vortex core; and \\
5) Hall anomaly and vortex many body effect. \\
For background material, the fundamental contributions went back
to Bardeen and Stephen \cite{bs} and to Nozieres and Vinen
\cite{nv}. The first direct measurement of the transverse force on
moving vortices in superconductors was reported in 1997 in Sweden
\cite{zhu}. More background materials can be found in cond-mat
\cite{ao03,ao04,ao05}.

It should be kept in mind here that we are talking about the
"mistakes" made by a very capable group of theoretical physicists.
Normally it is already rare to make one such "mistake" by any of
them. Instead, five have been identified collectively. To some, it
would be possible that there could exist mistake(s) in the
identification process. Thus, it is reasonable to ask how it goes
after two years.

A survey of recent literature has shown that till now three such
mistakes have been actually confirmed during past two years by
other researchers. Given the amount of intellectual effort, this
is a rather impressive number: 3 out 5, already a majority. In the
following we discuss them one by one.

\section{Misuse of the relaxation time approximation}

The most subtle inconsistency is the relaxation time approximation
employed by Kopnin and his co-workers (Kopnin and Kratsov, JETP
Lett.,1976; Kopnin and Lopatin, Phys. Rev. B, 1995; van Otterlo,
Feigelman, Geshkenbein, Blatter, Phys. Rev. Lett., 1995, see the
discussions and references in Ref's. \cite{ao04,ao05}. This type
of mistakes frequently occurs in the force-balance type
calculation of transport coefficients, already noticed at least by
Green in the 1940`s and extensively and repeatedly discussed by R.
Kubo. Unfortunately, as R. Kubo remarked in his coauthored famous
book on statistical physics, in the literature such error
repeatedly appears in different disguises.

Expounding a statement in Thouless, Ao, and Niu \cite{tan}, it was
explicitly formulated that vortex friction can be obtained without
such relaxation time approximation \cite{az}. The Bardeen-Stephen
case with disorder for s-wave superconductors was then explicitly
verified within this formulation \cite{az}. There is no influence
on Magnus force, expected from Anderson's dirty superconductor
theorem and the topology (Berry phase, spectral flow, chiral
symmetry, etc).

Now, this formulation \cite{az} has been further extended to the
d-wave superconductors \cite{nikolic}. From the point of vortex
core contribution to vortex friction, this is the first nontrivial
result in more than 40 years since Bardeen-Stephen \cite{bs}.

\section{Mixing with another effect}

The second type of mistake is very interesting. It concerns the
contribution of "phonons" (Sonin, Soviet Phys. JETP,1976; Phys.
Rev. B, 1997, see the discussions and references in Ref's.
\cite{ao04,ao05}).

The problem is clear: At zero temperature according to Sonin there
should not exist any effect. According to a group of researchers
loosely associated with Seattle, there should be an dissipative
effect at zero temperature: Zero for constant vortex velocity but
finite when vortex accelerating \cite{nat,dan96,la}. This is a
non-conventional damping, corresponding to the non-ohmic damping
in the quantum dissipative dynamics studied by Leggett.

Now, such effect has been suggested to be responsible to account
for the dissipation puzzle in quantum turbulence at zero
temperature \cite{barenghi}, closely resembles to the process
discussed by Vinen \cite{vinen}.

\section{Hall Anomaly and Vortex Many Body Effect}

This is the so-called Hall anomaly: How could the large and
universal transverse force (Magnus force) lead to a small Hall
angle, and, sometime to a sign change? The ingredients to explain
the Hall anomaly are actually already implied in the works of
Dirac and Abrikosov.

Let us stand one step away from vortex dynamics and consider the
Hall effect in semiconductors.  The transverse force there, the
Lorentz force, on a moving electron is universal. There exist such
extremely rich Hall phenomena in semiconductors: small Hall angle,
sign change, Quantum Hall effect, etc. It would be puzzling that
how could a universal Lorentz force generate such a complexity.
Fortunately, this puzzle has long been solved: The competition
between electron many body effect (Coulomb interaction,
Fermi-Dirac statistics) and pinning (lattice, impurities, etc).
The key to solve the puzzle is logically parallel to Dirac`s idea
of a void in a filled Fermi sea: the existence of holes in a
filled energy band.

The ubiquitous existence of the Abrikosov lattice, the starting
point of theoretical considerations in the mixed state of
superconductors, already loudly suggests that one must consider
vortex many body effect. Following this suggestion, it was argued
in 1995 that the competition between vortex many body interaction
and pinning can explain the Hall anomaly in the mixed state
\cite{ao95}. It may be further pointed out that the vortex many
body effect-pinning model appears consistent with all major
experimental observations on Hall anomaly \cite{ao98a}. The
earlier theories by Kopnin and his associates/collaborators on
Hall anomaly based on independent vortex dynamics model have been
shown to be wrong both mathematically \cite{ao98b} and physically
\cite{ao99}.

Because of those critiques on their work, Kopnin and Vinokur later
published a paper \cite{kv} completely embraced the idea of vortex
many body plus pinning for Hall anomaly. The interesting thing is
that they completely "forgot" the relevant prior theoretical work.
It may be instructive to compare the abstract of Ao, 1998
\cite{ao98a}: \\
{\it Physical arguments are presented to show that the Hall
anomaly is an effect of the vortex many-body correlation rather
than that of an individual vortex. Quantitatively, the
characteristic energy scale in the problem, the vortex vacancy
formation energy, is obtained for thin films. At low temperatures
a scaling relation between the Hall and longitudinal resistivities
is found, with the power depending on sample details. Near the
superconducting transition temperature and for small magnetic
fields the Hall conductivity is found to be proportional to the
inverse of the magnetic field and to the quadratic of the
difference between the measured and the transition temperatures.}
\\
to the abstract of Kopnin and Vinokur, 1999 \cite{kv}: \\
{\it We demonstrate that pinning strongly renormalizes both
longitudinal and Hall resistivity in the flux flow regime. Using a
simple model for the pinning potential we show that the magnitude
of the vortex contribution to the Hall voltage decreases with
increase in the pinning strength. The Hall resistivity rho(xy)
scales as rho(xx)(2) only for a weak pinning. On the contrary, a
strong pinning breaks the scaling relation and can even result in
a sign reversal of rho(xy).} \\
The readers are advised to read the full original articles in the
span of 10 years. It is clear that once the correct physical
understanding has been reached, there exists numerical other
mathematical approaches.

During past two years, explicit and controllable experiment was
designed to test the idea of vortex many body effect for Hall
anomaly. The experimental results were surprisingly consistent
with theoretical predictions \cite{ghenim}.

We also witness further development of theory based on this vortex
many-body effect plus pinning for Hall anomaly from the competing
research group \cite{shklovskij}.

Because same results have been repeatedly obtained by competing
groups it may be concluded that the Hall anomaly is now solved at
least conceptually. This is not an easy achievement given its
vortex-like history bothering at least two generations of
condensed matter physicists.

\section{Further progress}

Though 3 out of 5 have been confirmed by other researchers, it is
still possible in the remaining 2 "mistakes" the present author
identified could be mistakes in themselves. Because nobody can
prediction the future, we have to wait for the further progresses
on vortex dynamics. The only thing we can hope is that this would
be too long. In the same time, it is reasonable to expect that the
three mistakes already identified will continue to be identified
in one form or another by more researchers.

{\ }

{\it The reference list here is incomplete. A balance reference
list was attempted in \cite{ao05}. Any suggestion on interesting
references will be appreciated. }

{\ }

\end{document}